\documentclass[%
 reprint, 
 amsmath,amssymb,
 aps, physrev,
floatfix, 
]{revtex4-2}

\usepackage{graphicx}
\usepackage{dcolumn}
\usepackage{bm}

\usepackage{xcolor}
\usepackage{xurl}
\usepackage{hyperref}
\usepackage{booktabs}
\usepackage[utf8]{inputenc}


\newcommand{\RNum}[1]{\uppercase\expandafter{\romannumeral #1\relax}}

\makeatletter
\newcommand{\figurecaptiontype}{\def\@captype{figure}}
\makeatother
\ExplSyntaxOn
\NewDocumentCommand{\layoutbiburl}{m}{
  \group_begin:
  \urlstyle{same}
  \tl_set:Nn \l_tmpa_tl {#1}
  \tl_replace_all:Nnn \l_tmpa_tl {\_} {_}
  \exp_args:NV \url \l_tmpa_tl
  \group_end:
}
\ExplSyntaxOff
\begin{document}

\preprint{APS/123-QED}

\title{\textbf{Chemical-space completeness through iterative crystal-structure generation and model adaptation}
}%

\author{Fengyu Xie}
\affiliation{
 College of Artificial Intelligence and Data Science, Suzhou Institute for Advanced Research, University of Science and Technology of China, Suzhou, Jiangsu 215123, China
}%

\author{Ruoyu Wang}
\affiliation{
 Suzhou National Laboratory, Suzhou, Jiangsu 215123, China
}%

\author{Taoyuze Lv}
\affiliation{
 College of Artificial Intelligence and Data Science, Suzhou Institute for Advanced Research, University of Science and Technology of China, Suzhou, Jiangsu 215123, China
}%

\author{Yuxiang Gao}
\affiliation{
 College of Artificial Intelligence and Data Science, Suzhou Institute for Advanced Research, University of Science and Technology of China, Suzhou, Jiangsu 215123, China
}%

\author{Hongyu Wu}
\affiliation{
 Suzhou National Laboratory, Suzhou, Jiangsu 215123, China
}%

\author{Zhicheng Zhong}
\email{zczhong@ustc.edu.cn}
\affiliation{
 College of Artificial Intelligence and Data Science, Suzhou Institute for Advanced Research, University of Science and Technology of China, Suzhou, Jiangsu 215123, China
}
\affiliation{
 Suzhou National Laboratory, Suzhou, Jiangsu 215123, China
}

\date{\today}

\begin{abstract}
The emergence of deep learning has brought large-scale exploration of crystalline materials closer to practical realization. Yet, universal atomistic models face a practical trade-off between chemical generality and the accuracy, efficiency, and adaptability required for intensive exploration of specific materials systems. Within a bounded chemical system, this trade-off can be relaxed by exploiting its limited chemical complexity. Guided by this intuition, we propose a chemical-system–centric strategy that couples crystal-structure generative models with machine-learned force fields (MLFFs) in an iterative generation–evaluation–refinement loop. Using Li--P--S as a test case, we generate approximately 70,000 candidate structures, including more than 10,000 stable–unique–novel structures. The diversity of near-equilibrium local environments saturates within the first few iterations, accompanied by convergence of MLFF prediction errors, providing an operational measure of chemical-space completeness within bounded systems. The exploration also recovers chemically plausible P–S motifs that are absent from the pretraining databases but supported by earlier experiments. The resulting system-adapted models and structures further enable finite-$P$–$T$ phase-stability calculations, Li-ion transport screening, and electronic-structure prediction. These results suggest that chemical-system–centric exploration provides a practical route toward data-efficient and high-fidelity modeling within bounded chemical spaces.
\end{abstract}

\maketitle


\section{Introduction}\label{sec: intro}

Exploring crystalline materials across broad composition and structural spaces remains a central challenge in computational materials science. A long-standing goal is to achieve accurate and efficient prediction of stable and metastable structures, their energetics, and associated properties across diverse chemical systems. Such capability would not only accelerate materials discovery, but also enable computational models to move from passive evaluation of predefined structures toward active exploration of previously unknown regions of materials space.

Recent advances in deep generative models for crystals (e.g., MatterGen\cite{zeniGenerativeModelInorganic2025}, DiffCSP\cite{jiaoCrystalStructurePrediction2023}, ADiT\cite{joshiAllatomDiffusionTransformers2025}, Uni3DAR\cite{luUni3DARUnified3D2025}) and machine-learned force fields (MLFFs) (e.g., DPA-3\cite{zhangGraphNeuralNetwork2025}, MatterSim\cite{yangMatterSimDeepLearning2024}, MACE\cite{batatiaMACEHigherOrder2022}, CHGNet\cite{dengCHGNetPretrainedUniversal2023}) have brought this objective closer to practical realization. Generative models explore latent structural spaces, allowing direct crystal-structure sampling beyond template filling\cite{schmidtDataset175kStable2022} or elemental substitution approaches\cite{hautierDataMinedIonic2011, merchantScalingDeepLearning2023}, and offering composition variability compared to traditional crystal-structure prediction methods\cite{pickardInitioRandomStructure2011a, oganovCrystalStructurePrediction2006, glassUSPEXEvolutionaryCrystal2006, wangCrystalStructurePrediction2010, wangCALYPSOMethodCrystal2012}. Meanwhile, MLFFs provide near-DFT accuracy in energy and force evaluation at orders-of-magnitude lower computational cost. In principle, coupling these two capabilities makes it possible to construct iterative discovery workflows in which new structures are generated, evaluated, selectively validated by density-functional theory (DFT), and used to refine subsequent exploration. This creates a route toward increasingly autonomous and data-efficient exploration of crystalline materials.

However, a key difficulty remains in balancing chemical universality with the accuracy, efficiency, and adaptability required for intensive exploration of a specific materials system. Large foundational models such as UMA\cite{woodUMAFamilyUniversal2025} exhibit impressive cross-chemistry transferability, but uniformly high accuracy across arbitrary compositions and underrepresented bonding environments cannot be assumed. Their broad coverage also requires greater model capacity and computational resources. In contrast, smaller and more adaptable models such as MatterSim, DPA-3, and MACE are easier to fine-tune and deploy efficiently within restricted chemistries, but may sacrifice universality. This trade-off motivates a complementary strategy. Rather than pursuing all-element generality, one may exploit the limited chemical complexity within a bounded elemental system to achieve high-accuracy, data-efficient exploration.

\begin{figure*}
    \centering
    \includegraphics[width=1.0\linewidth]{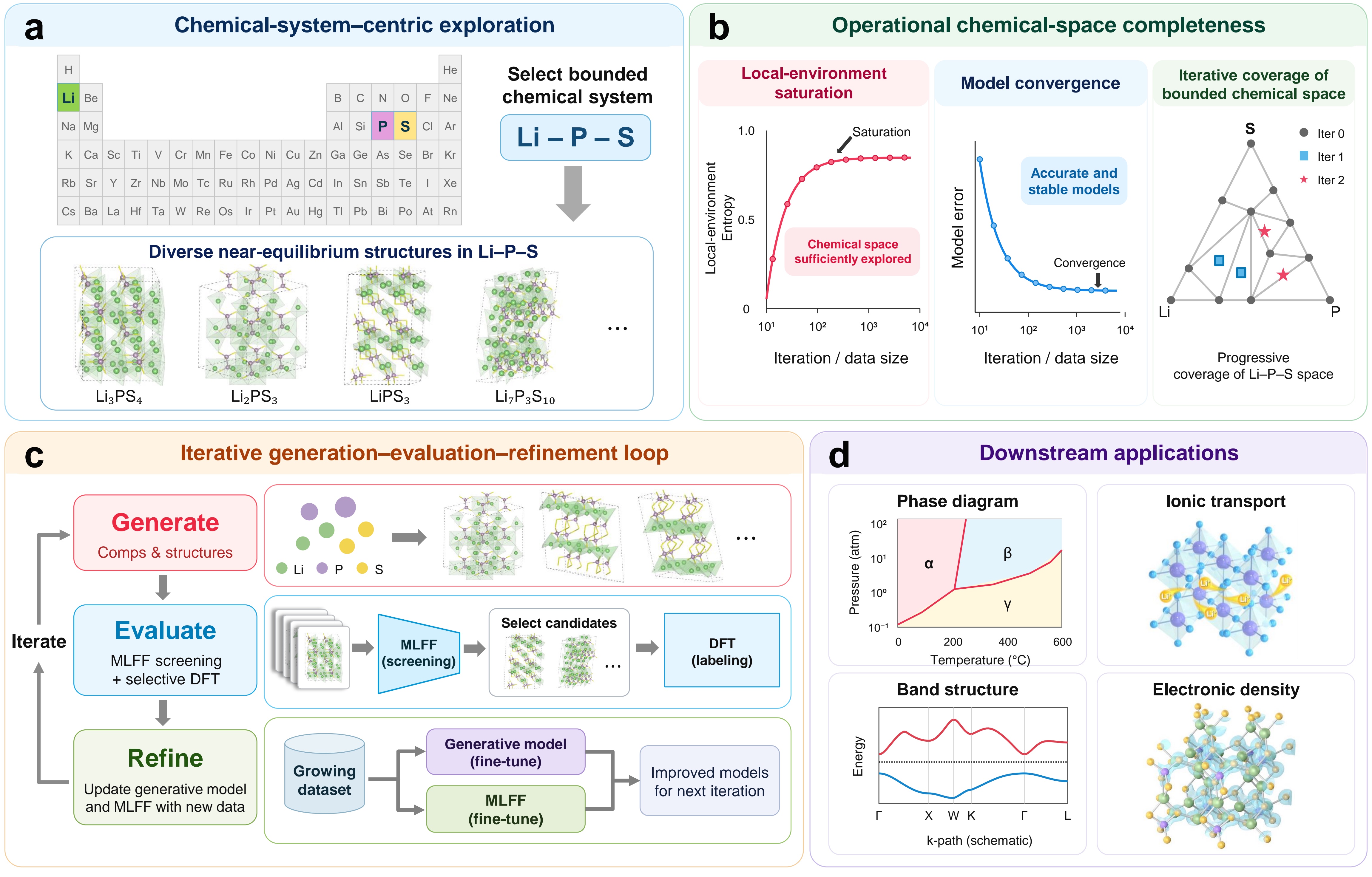}
    \caption{Schematic illustration of the chemical-system–centric exploration framework and its downstream applications. (a) A bounded chemical system, Li--P--S, is first selected from the full elemental space. Representative near-equilibrium structures within this chemical space are then explored. (b) Operational chemical-space completeness is assessed using three indicators: saturation of local chemical environments, convergence of model error, and progressive coverage of the bounded composition space. (c) Iterative generation–evaluation–refinement loop. Generative models propose candidate crystal structures; MLFFs perform rapid screening of candidates for selective DFT calculations; the resulting data are incorporated into a growing dataset to refine both the generative model and the MLFF for the next iteration. (d) Once adapted to the target chemical system, the resulting models and structures support downstream calculations within the same chemical space, including phase-diagram construction, ionic-transport screening, band-structure analysis, and electronic-density prediction.}
    \label{fig:fig1}
\end{figure*}

The nature of chemical bonding provides an important physical intuition. As suggested by the Materials Project\cite{hortonAcceleratedDatadrivenMaterials2025}, most experimentally relevant crystalline materials can be organized into chemical systems comprising a small ($<7$) and finite set of elements. Within such a bounded system, the diversity of short-range bonding types and local geometries is constrained by valence, coordination, and local chemical compatibility. For instance, carbon predominantly adopts $sp^3$, $sp^2$, or $sp^1$ hybridization even in amorphous networks \cite{robertsonAmorphousCarbon1986, robertsonElectronicAtomicStructure1987}, while silicon in silicates typically forms SiO$_4$ tetrahedra \cite{liebauStructuralChemistrySilicates1985}. Once such local building blocks are covered, much of the remaining structural diversity arises from their spatial arrangements and recombination. These observations suggest that a bounded chemical system may exhibit an operational notion of completeness over near-equilibrium configurations and short-range atomic environments, without requiring exhaustive coverage of the full configurational landscape. Grounded in this intuition, we introduce a chemical-system–centric strategy for crystalline materials exploration. Here, “chemical-space completeness” refers specifically to the operational saturation of short-range atomic environments associated with near-equilibrium stable and metastable crystalline configurations. We probe this regime through an iterative generation–evaluation–refinement loop that couples generative models and MLFFs: the generative model proposes candidate structures, the MLFF evaluates their energetics, and selected DFT data are used to refine both models (see Figure~\ref{fig:fig1}).

We demonstrate this strategy on the Li--P--S ternary system, a prototypical yet chemically intricate solid-state electrolyte chemistry\cite{sakudaSulfideSolidElectrolyte2013, zhangSulfideBasedSolidStateElectrolytes2019}. Phosphorus and sulfur form diverse bonding networks, including tetrahedral PS$_4$ units, P–P–linked trigonal motifs, and bridging or polymeric sulfide species, producing complex anionic frameworks coupled to lithium coordination. This makes Li--P--S a stringent test case for whether short-range chemical environments within a bounded system can become effectively saturated despite substantial structural diversity. Over seven exploration cycles, our framework generated approximately 70,000 candidate structures, more than 10,000 of which satisfy the stable–unique–novel criterion\cite{zeniGenerativeModelInorganic2025}. The diversity of local environments sampled by the iterative exploration saturated rapidly within two to three iterations, while the MLFF concurrently converged to meV-level accuracy in energy and force with a limited number of DFT-labeled configurations. Notably, although the generative model was pretrained exclusively on structures from the Materials Project\cite{jainCommentaryMaterialsProject2013} and Alexandria databases\cite{schmidtDataset175kStable2022, schmidtMachineLearningAssistedDeterminationGlobal2023}, the exploration autonomously recovered a broad range of chemically plausible P–S anionic motifs and metastable phases absent from those databases. These include the (PS$_3$)$_3^{3-}$ trimer, the P$_3$S$_8^{3-}$ ring, two isomers of the P$_2$S$_8^{2-}$ ring, and polymeric (PS$_4$)$_n^{n-}$ chains. These motifs and their characteristic P–P and S–S linkages are consistent with earlier experimental reports, suggesting that generative exploration constrained by MLFF-based energetic evaluation can, to some extent, recover and generalize chemically plausible bonding patterns beyond the structures explicitly represented in the training data.

Beyond structural generation, the resulting chemical-system-adapted models support a range of downstream calculations within the same chemical space, including finite-temperature and finite-pressure phase-diagram construction, ionic-transport screening, and electronic-structure prediction. In principle, this iterative generation and fine-tuning strategy can be extended to other multicomponent chemical systems, providing a scalable route to high-fidelity materials exploration without requiring universal coverage of the full elemental space.

\section{Methods}\label{sec:method}

\subsection{Generative model}
We primarily employed MatterGen\cite{zeniGenerativeModelInorganic2025} as the generative backbone due to its built-in support for complex conditional generation, while other crystal generation models such as Con-CDVAE\cite{yeConCDVAEMethodConditional2024}, ADiT\cite{joshiAllatomDiffusionTransformers2025}, and Uni3DAR\cite{luUni3DARUnified3D2025} can in principle be adapted within the same iterative exploration scheme.

Starting from the open-source, $E_\mathrm{hull}$- and chemical-system-conditional version \cite{MattergenCheckpointsChemical_system_energy_above_hull} of MatterGen, we generated 9,984 candidate structures per iteration—3,328 for each conditioned $E_{\mathrm{hull}}$ (0.00, 0.03, and 0.06 eV/atom). Preliminary relaxations were conducted using the MLFF, and $E_{\mathrm{hull}}$ values were evaluated relative to the MP database and previously generated structures.

S.U.N. structures were selected following the same criteria introduced in the original MatterGen work:
\begin{itemize}
    \item Stability: $E_\mathrm{hull} < 0.1$ eV/atom.
    \item Uniqueness: no identical structure within the current iteration, identified by pymatgen's~\cite{ongPythonMaterialsGenomics2013} StructureMatcher.
    \item Novelty: not present in earlier iterations or in the MP database, also verified using StructureMatcher.
\end{itemize}

After DFT evaluation, the generative model was fine-tuned using DFT-computed $E_\mathrm{hull}$ labels to bias subsequent generations toward lower-energy configurations.

\subsection{DFT computations and dataset preparation}
In each iteration, up to 600 MLFF-relaxed structures were selected for DFT labeling according to $E_\mathrm{hull}$ intervals: $\leq$100 for $E_\mathrm{hull} < 0$, 200 for $0\leq E_\mathrm{hull}<0.03$, and 300 for $0.03\leq E_\mathrm{hull}<0.10$ eV/atom. Typically, 400–600 structures were sampled. Each structure was randomly perturbed three times in atomic coordinates and lattice parameters before DFT relaxation. From each trajectory, 1–5 ionic steps, including the initial and final states, were randomly extracted.

To enrich the dataset, five additional static calculations were performed on random perturbations of each converged structure, producing a set of compositionally and structurally diverse configurations with corresponding energies and forces. Each iteration yielded approximately 10,000 data points. The perturbation amplitudes followed normal distributions of 0.08–0.15 $a$ for atomic positions (where $a = (V_\mathrm{cell}/N_\mathrm{atom})^{1/3}$) and 3--5 \% for lattice constants.

All DFT calculations, including structural relaxations and static perturbations, were carried out using VASP with the projector-augmented-wave (PAW) method\cite{kresseEfficiencyAbinitioTotal1996, kresseEfficientIterativeSchemes1996, kresseUltrasoftPseudopotentialsProjector1999} and the Perdew-Burke-Ernzerhof (PBE) exchange–correlation functional\cite{perdewGeneralizedGradientApproximation1996, perdewGeneralizedGradientApproximation1997}. Input files were generated using pymatgen’s MPRelaxSet with default parameters. A plane-wave cutoff of 520 eV and a $\Gamma$-centered k-point mesh with a reciprocal-space density of 64 $\mathrm{\AA}^3$ were used. Li 1s electrons were treated as valence electrons. Electronic convergence was set to $5 \times 10^{-5}$ eV/atom, and ionic steps were converged to 10 times the electronic threshold.

To ensure internal consistency, all Li--P--S structures queried from the MP database (referred to as iteration 0) were recalculated under identical DFT settings.

\subsection{MLFFs}
We benchmarked our framework using three MLFF architectures of comparable million-parameter size: DPA 3.1-3M\cite{Models343DPA313M}, MACE-MPA-0-medium\cite{ACEsuitMace2025}, and MatterSim-5M\cite{MatterSimPretrained_modelsMain}. The neighbor cutoff radii were set to $6 \AA$ for DPA-3 and MatterSim, and $5 \AA$ for MACE. Each model was fine-tuned using architecture-specific training settings chosen to ensure stable convergence rather than an identical epoch budget.

For DPA-3, models were trained for approximately 50 epochs (corresponding to $\sim50\times N_{\mathrm{train}}$ steps) with automatic batch sizing. The learning rate started at 0.001 and decayed exponentially every 2000 steps to $3\times10^{-5}$. The total loss combined energy, force and virial terms with gradually adjusted mixing ratio, starting at 0.2:100:0.02 and ending at 20:60:1.

For MatterSim, training was performed for 200 epochs, with an initial learning rate of 0.001 that decayed by a factor 0.8 every 4 epochs.

For MACE, models were first trained for 400 epochs under the exponential moving average (EMA) mode, starting at a learning rate of 0.01 that decayed by 0.8 every 20 epochs once the validation plateaued followed by 100 in the stochastic weight averaging (SWA) mode at a fixed learning rate of 0.001. The validation fraction was set to 5\%.

In each iteration, training structures were selected according to the maximum-information-entropy-gain criterion. Following Reference~\cite{schwalbe-kodaModelfreeEstimationCompleteness2025}, information entropy of a set of $n$ structures ($H$) is defined as
\begin{equation}
    H(\mathbf{\{X\}}) = - \frac{1}{n} \sum_{i=1}^{n} \log \left[\frac{1}{n} \sum_{j=1}^{n} K_h(\mathbf{X}_i, \mathbf{X}_j) \right]
\end{equation}
where $\mathbf{X}_i$ denotes the vectorized local-environment descriptors of structure $i$, $K_h$ is a Gaussian kernel, and $h$ is a tunable bandwidth (default value used). The entropy gain score used for selecting a new data point $\mathbf{Y}$ into the existing dataset $\mathbf{\{X\}}$ is defined as
\begin{equation}
    \delta H(\mathbf{Y}|\mathbf{\{X\}}) = -\log \left[\sum_{i=1}^{n} K_h(\mathbf{Y}, \mathbf{X}_i)\right].
\end{equation}
All DFT-labeled structures were ranked by $\delta H$, and the top $N_{\mathrm{train}}$ configurations were selected to enrich the information diversity of the training set while minimizing computational cost. Training data were drawn only from iterations 0 to 3, whereas the unselected data from these iterations, along with all data from subsequent iterations, were reserved for testing. Based on the benchmark results described later in Section~\ref{sec:ft}, we used the DPA-3 model fine-tuned with $N_\mathrm{train}=4050$ for subsequent stability estimation and phase-diagram computations. The same entropy measure was also used to monitor the growth of local-environment diversity during iterative exploration, with the diminishing entropy gain $dH/dN$ serving as an operational indicator of saturation.

Overall, the entire exploration required approximately 48,000 CPU core-hours on 32-core Xeon Gold nodes and 800 GPU hours on NVIDIA V100 GPUs.

\section{Results}\label{sec: res}
\subsection{Evolution of the generated stability distribution}

\begin{figure*}
    \centering
    \includegraphics[width=1.0\linewidth]{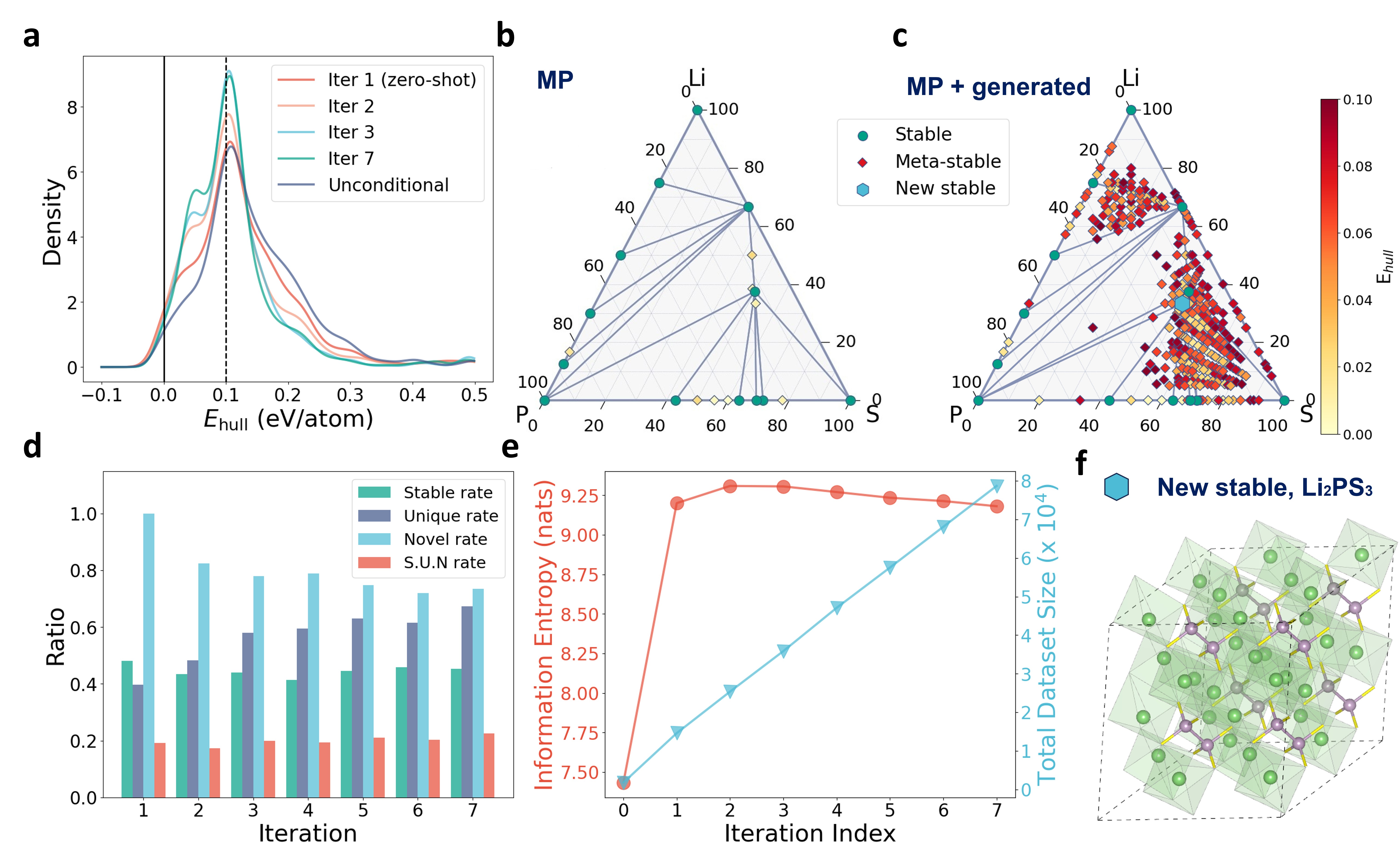}
    \caption{(a) Density distribution of generated structures over $E_{\mathrm{hull}}$ within the Li--P--S chemical space. (b–c) Zero-temperature phase diagrams including all Li--P--S phases from (b) the Materials Project (MP) database and (c) the MP database combined with all iteratively generated S.U.N. structures. Green circles denote stable phases on the convex hull, while the blue hexagon marks the newly discovered Li$_2$PS$_3$ stable phase absent from MP. Yellow-to-red dots indicate metastable structures with $E_{\mathrm{hull}} \leq 0.1$ eV/atom, where darker red corresponds to higher $E_\mathrm{hull}$ and thus lower stability. (d) Ratios of stable (green), unique (blue), novel (cyan), and total S.U.N. (red) structures as a function of iteration number. (e) Information entropy of local-environment descriptors (red dots) and the total number of DFT calculations (cyan triangles) versus iteration number. Iteration 0 corresponds to Li--P--S structures originally included in the MP database. (f) Crystal structure of the newly generated stable Li$_2$PS$_3$ phase, highlighted as a blue hexagon in (c).}
    \label{fig:fig2}
\end{figure*}

We first examine how energetic conditioning and iterative refinement reshape the stability distribution of generated structures. As shown in Figure~\ref{fig:fig2}a, unconditional generation constrained only by element types produces a broad, unimodal $E_{\mathrm{hull}}$ distribution spanning approximately -0.05 to 0.4 eV/atom. In contrast, generation conditioned on $E_{\mathrm{hull}}=0.0$ eV/atom yields a narrower distribution with higher weight at lower energies. Subsequent fine-tuning using DFT-computed $E_{\mathrm{hull}}$ labels further increases the fraction of low-energy structures over successive iterations, indicating that the generative model progressively adapts toward thermodynamically more favorable configurations.
During this exploration, numerous metastable structures absent from the Materials Project (MP)\cite{jainCommentaryMaterialsProject2013, hortonAcceleratedDatadrivenMaterials2025} were generated. Notably, the workflow also recovered a stable Li$_2$PS$_3$ phase containing P--P-bonded P$_2$S$_6^{4-}$ units, which is absent from the MP database but has been reported experimentally\cite{mercierSyntheseStructureCristalline1982, dietrichLocalStructuralInvestigations2016} (Figure~\ref{fig:fig2}b,c). These results show that iterative energetic refinement shifts the generated distribution toward the near-equilibrium region while continuing to sample metastable structures beyond existing database entries.

\subsection{Operational saturation of local chemical environments}\label{sec:saturation}

Despite the continued generation of stable–unique–novel (S.U.N.)\cite{zeniGenerativeModelInorganic2025} structures across iterations (Figure~\ref{fig:fig2}d, in which the remaining S.U.N. fraction is approximately 18\%), the information entropy\cite{schwalbe-kodaModelfreeEstimationCompleteness2025} of local-environment descriptors rapidly approaches saturation within the first two to three cycles and subsequently changes only slightly (Figure~\ref{fig:fig2}e). Consistently, Appendix Figure~\ref{fig:figs2} shows that the distributions of local structural fingerprints\cite{zimmermannLocalStructureOrder2020}, projected using t-SNE\cite{maatenVisualizingDataUsing2008}, overlap strongly after two iterations.
These observations indicate that successive exploration contributes progressively less new short-range chemical information. Importantly, this local saturation does not imply loss of global structural diversity, as later iterations continue to generate S.U.N. structures through alternative arrangements and recombinations of previously sampled motifs. We use this diminishing gain in local-environment diversity as an operational indicator of chemical-space completeness within the near-equilibrium Li--P--S configurations considered here.

\subsection{Convergence of machine-learned force fields}\label{sec:ft}
\begin{figure*}
    \centering
    \includegraphics[width=1.0\linewidth]{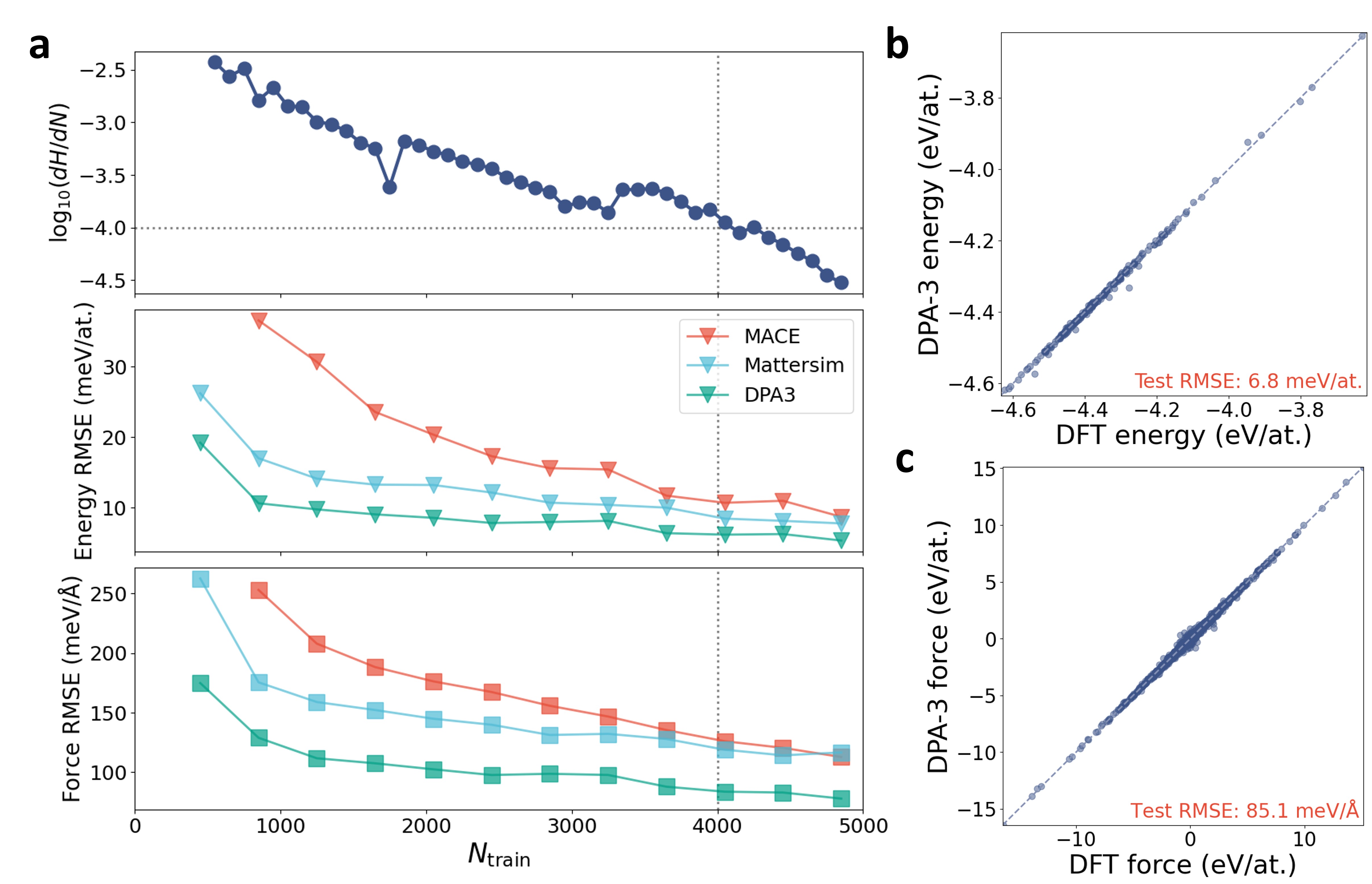}
    \caption{Fine-tuning performance of machine-learned force fields in the Li--P--S chemical system. (a) Logarithm of the information-entropy increase rate ($\mathrm{log}_{10}(dH/dN)$, upper panel), tested energy RMSE (middle panel), and force RMSE (lower panel) as functions of the number of training frames ($N_{\mathrm{train}}$). (b–c) Parity plots comparing DFT-calculated and DPA-3-predicted (b) energies and (c) forces on the test set. The DPA-3 model shown here is fine-tuned using $N_{\mathrm{train}}=4050$.}
    \label{fig:fig3}
\end{figure*}

As the diversity of local chemical environments approaches saturation, the MLFFs show a corresponding convergence in predictive accuracy. Using configurations selected by the maximum-information-entropy-gain criterion (see Methods), we fine-tuned DPA-3\cite{zhangGraphNeuralNetwork2025}, MACE\cite{batatiaMACEHigherOrder2022}, and MatterSim\cite{yangMatterSimDeepLearning2024} on progressively larger training sets and evaluated their performance on held-out Li--P--S configurations.
As shown in Figure~\ref{fig:fig3}a, the decrease in information-entropy gain is accompanied by a systematic reduction in both energy and force errors. Among the compared models, DPA-3 consistently provides the lowest test errors. At $N_\mathrm{train}=4050$, where $dH/dN$ falls below $10^{-4}$ per structure, the fine-tuned DPA-3 model reaches an energy RMSE of 6.8 meV/atom and a force RMSE of 85.1 meV/$\mathrm{\AA}$ (Figure~\ref{fig:fig3}b,c). The corresponding RMSE in $E_\mathrm{hull}$ decreases from 46.9 to 26.5 meV/atom (Appendix Figure~\ref{fig:figs3}).
The improvement becomes substantially smaller beyond approximately 1000 training configurations. At $N_\mathrm{train}=1250$, the energy and force RMSEs are already 9.8 meV/atom and 111.5 meV/$\mathrm{\AA}$, respectively, sufficient for large-scale stability screening and ranking. These results indicate that the reduction in new local chemical information is accompanied by diminishing returns in model improvement, providing a second, complementary indicator of convergence within the explored Li--P--S domain.

\subsection{Generative recovery of chemically plausible motifs beyond the training data}

\begin{figure*}
    \centering
    \includegraphics[width=0.9\linewidth]{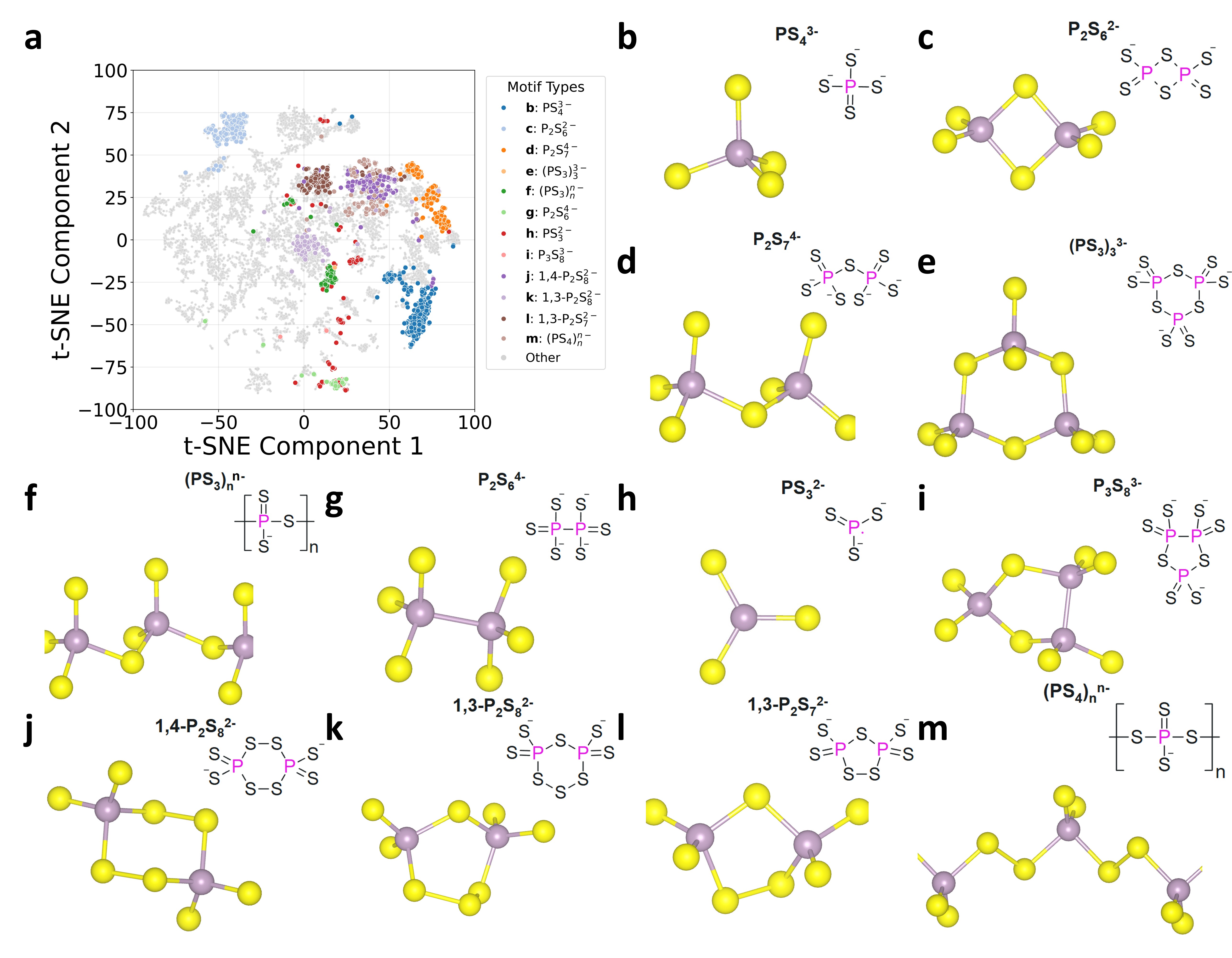}
    \caption{Generative diversity of P–S anionic motifs in the Li--P--S chemical space. (a) Distribution of the first two t-SNE components derived from the local structural fingerprints of all generated structures (excluding Li atoms). Colored clusters highlight groups of structures containing representative anion motifs as shown in b--m. (b--m) Representative thiophosphate anions discovered in the iterative process. Purple and yellow spheres represent P and S atoms, respectively. (b) PS$_4^{3-}$ tetrahedron anion. (c) P$_2$S$_6^{2-}$ edge-sharing anion. (d) P$_2$S$_7^{4-}$ corner-sharing anion. (e) P$_3$S$_9^{3-}$ corner-sharing ring anion. (f) (PS$_3$)$_n^{n-}$ polymeric chain. (g) P$_2$S$_6^{4-}$ dimer with P-P bond. (h) PS$_3^{2-}$ triangular monomer. (i) P$_3$S$_8^{3-}$ ring anion. (j) 1,4-P$_2$S$_8^{2-}$, with two P atoms occupying para positions (1,4-substitution) within a six-membered P$_2$S$_4$ ring, lying opposite to each other across two S-S bridges. (k) 1,3-P$_2$S$_8^{2-}$, with two P atoms occupying meta positions (1,3-substitution) within a six-membered P$_2$S$_4$ ring. (l) 1,3-P$_2$S$_7^{2-}$, featuring two P atoms in meta positions within a five-membered P$_2$S$_3$ ring. (m) (PS$_4$)$_n^{n-}$ polymeric chain linked by S-S bonds.}
    \label{fig:fig4}
\end{figure*}

To examine whether the iterative exploration generalizes beyond the structures explicitly represented in the training data, we analyzed the P--S anionic motifs emerging across the generated Li--P--S structures. Figure~\ref{fig:fig4}a shows the distribution of local structural fingerprints of all S.U.N. structures (excluding Li atoms), projected using t-SNE, with representative thiophosphate motifs highlighted in Figure~\ref{fig:fig4}b–m.
Relative to the MP2020 + Alexandria\cite{schmidtDataset175kStable2022, schmidtMachineLearningAssistedDeterminationGlobal2023} dataset (Alex–MP), on which the generative model was pretrained\cite{zeniGenerativeModelInorganic2025}, the exploration reproduces a broad range of known anionic units. These include motifs already present in Li--P--S compounds, such as the PS$_4^{3-}$ tetrahedron and (PS$_3$)$_n^{n-}$ chain, as well as motifs represented only in other thiophosphate chemistries, including P$_2$S$_6^{2-}$, P$_2$S$_7^{4-}$, PS$_3^{2-}$, and P$_2$S$_6^{4-}$ units.

More importantly, the generated structures also contain several ring-like and polymeric motifs absent from the combined Alex–MP training set, including (PS$_3$)$_3^{3-}$, P$_3$S$_8^{3-}$, two isomers of P$_2$S$_8^{2-}$, and polymeric (PS$_4$)$_n^{n-}$. Several of these motifs, together with their characteristic P--P and S--S linkages, are consistent with earlier experimental reports in thiophosphate chemistry. Their occurrence therefore suggests that generative exploration constrained by MLFF-based energetic evaluation does not merely reproduce known structural templates, but can recombine and extend learned local bonding patterns into chemically plausible configurations beyond the explicit training dataset. A summary of motif occurrences, their presence in the training data, and corresponding experimental references is provided in Table~\ref{tab:motifs}.

\begin{table*}[t]
\caption{Summary of representative P–S anion motifs identified in the generative exploration, their presence in the Alex–MP training dataset, and corresponding experimental references.}
\label{tab:motifs}
\begin{tabular}{@{}llll@{}}
\toprule
Formula & In Alex–MP, Li--P--S  & In Alex–MP, other systems & Experimental ref.\\
\midrule
PS$_4^{3-}$   & \checkmark (Li$_3$PS$_4$) & \checkmark (Na$_3$PS$_4$, K$_3$PS$_4$) & \cite{mercierStructureTetrathiophosphateLithium1982, hommaCrystalStructurePhase2011} \\
P$_2$S$_6^{2-}$ & $\times$ & \checkmark (KPS$_3$, CsPS$_3$, CuPS$_3$, etc.) & \cite{toffoliStructureCristallineLhexathiodimetaphosphate1978, brocknerChemInformAbstractCRYSTAL1985, hankoA2CuP3S91998, dietrichSynthesisStructuralCharacterization2017} \\
P$_2$S$_7^{4-}$ & $\times$ & \checkmark (LiCrP$_2$S$_7$, Zn$_2$P$_2$S$_7$, KVP$_2$S$_7$, etc.) & \cite{maierPrinciplesPhosphorusChemistry1962, yamaneCrystalStructureSuperionic2007} \\
(PS$_3$)$_3^{3-}$ & $\times$ & $\times$ & \cite{wolfChemInformAbstractCHEMISTRY1983, faliusCyclischeThiophosphateProdukte1992, hankoA2CuP3S91998} \\
(PS$_3$)$_n^{n-}$ & \checkmark (LiPS$_3$) & \checkmark (KPS$_3$, CuPS$_3$) & \cite{maierPrinciplesPhosphorusChemistry1962, eckertStructuralTransformationNonoxide1990, dietrichSpectroscopicCharacterizationLithium2018} \\
P$_2$S$_6^{4-}$ & $\times$ & \checkmark (LiAlP$_2$S$_6$, MnPS$_3$, MgPS$_3$, etc.) & \cite{mercierSyntheseStructureCristalline1982, dietrichLocalStructuralInvestigations2016} \\
PS$_3^{2-}$ & $\times$ & \checkmark (K$_2$CdP$_2$S$_6$, Cs$_3$NaP$_2$S$_6$, etc.) & \cite{mercierSyntheseStructureCristalline1982} \\
P$_3$S$_8^{3-}$ & $\times$ & $\times$ & \cite{faliusCyclischeThiophosphateProdukte1992} \\
1,4-P$_2$S$_8^{2-}$ & $\times$ & $\times$ & \cite{minshallPyridinium2255tetrathioCyclo1978, jonesTriethylammonium3366tetrathioxocyclodiphosphadithianate1781991, faliusCyclischeThiophosphateProdukte1992} \\
1,3-P$_2$S$_8^{2-}$ & $\times$ & $\times$ & \cite{gruberZweiNeueCyclische1997} \\
1,3-P$_2$S$_7^{2-}$ & $\times$ & $\times$ & \cite{jandaliDarstellungUndKristallstruktur1985, faliusCyclischeThiophosphateProdukte1992} \\
(PS$_4$)$_n^{n-}$ & $\times$ & $\times$ & \cite{katoLithiumionconductiveSulfidePolymer2021} \\
\bottomrule
\end{tabular}
\end{table*}

\subsection{Downstream applications: P–T phase stability}

The converged Li--P--S MLFF can be used to evaluate phase stability beyond zero-temperature conditions. Using the quasi-harmonic approximation\cite{togoFirstPrinciplesPhonon2015}, we computed finite-temperature and finite-pressure free energies for structures within the explored Li--P--S chemical space.

\begin{figure*}
    \centering
    \includegraphics[width=1.0\linewidth]{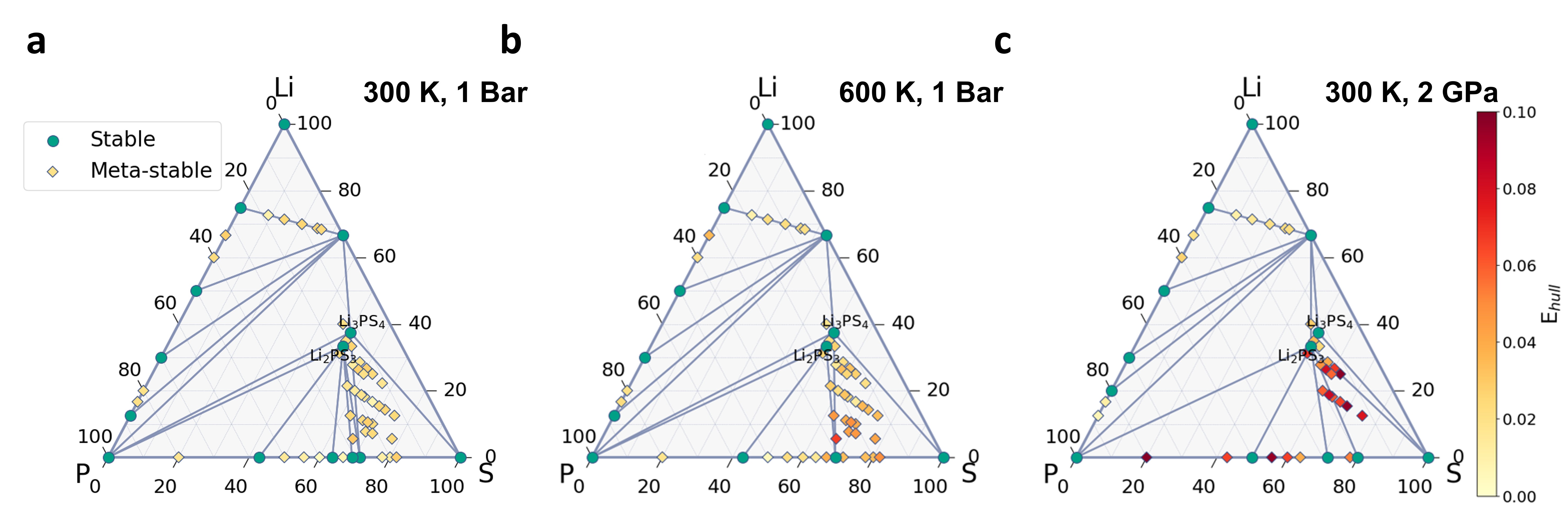}
    \caption{Finite-temperature, finite-pressure phase diagrams in Li--P--S chemical space at varied temperatures and pressures. (a) Phase diagram at T = 300 K, P = 1 bar. (b) Phase diagram at T = 600 K, P = 1 bar. (c) Phase diagram at $T = 300$ K, $P = 2$ GPa. Green dots represent stable phases on the convex hull, while triangles represent metastable phases with $E_{\mathrm{hull}} < 0.1$ eV/atom, and darker red color indicates lower stability.}
    \label{fig:fig5}
\end{figure*}

Figure~\ref{fig:fig5}a–c shows the resulting phase diagrams at representative thermodynamic conditions. Increasing the temperature from 300 to 600 K at 1 bar produces only modest changes in phase stability, with a slight increase in $E_{\mathrm{hull}}$ for low-Li compositions near the P–S boundary. In contrast, pressure has a stronger effect. At 2 GPa, S-rich phases above approximately 50 at.\% S become less stable, consistent with their lower atomic packing densities and larger \(PV\) contributions to the Gibbs free energy (Appendix Figure~\ref{fig:figs4}a).

Across the investigated range of 300–600 K and 1 bar–2 GPa, Li$_2$PS$_3$ and Li$_3$PS$_4$ remain thermodynamically stable. Their characteristic P$_2$S$_6^{4-}$ and PS$_4^{3-}$ anionic units are preserved, although their preferred spatial arrangements change with temperature and pressure (Appendix Figure~\ref{fig:figs4}b–e). These results illustrate that the system-adapted MLFF can be directly used for finite-$P$--$T$ thermodynamic investigations within the explored Li--P--S domain.

\subsection{Downstream applications: Ionic-transport screening}

\begin{figure*}
    \centering
    \includegraphics[width=0.9\linewidth]{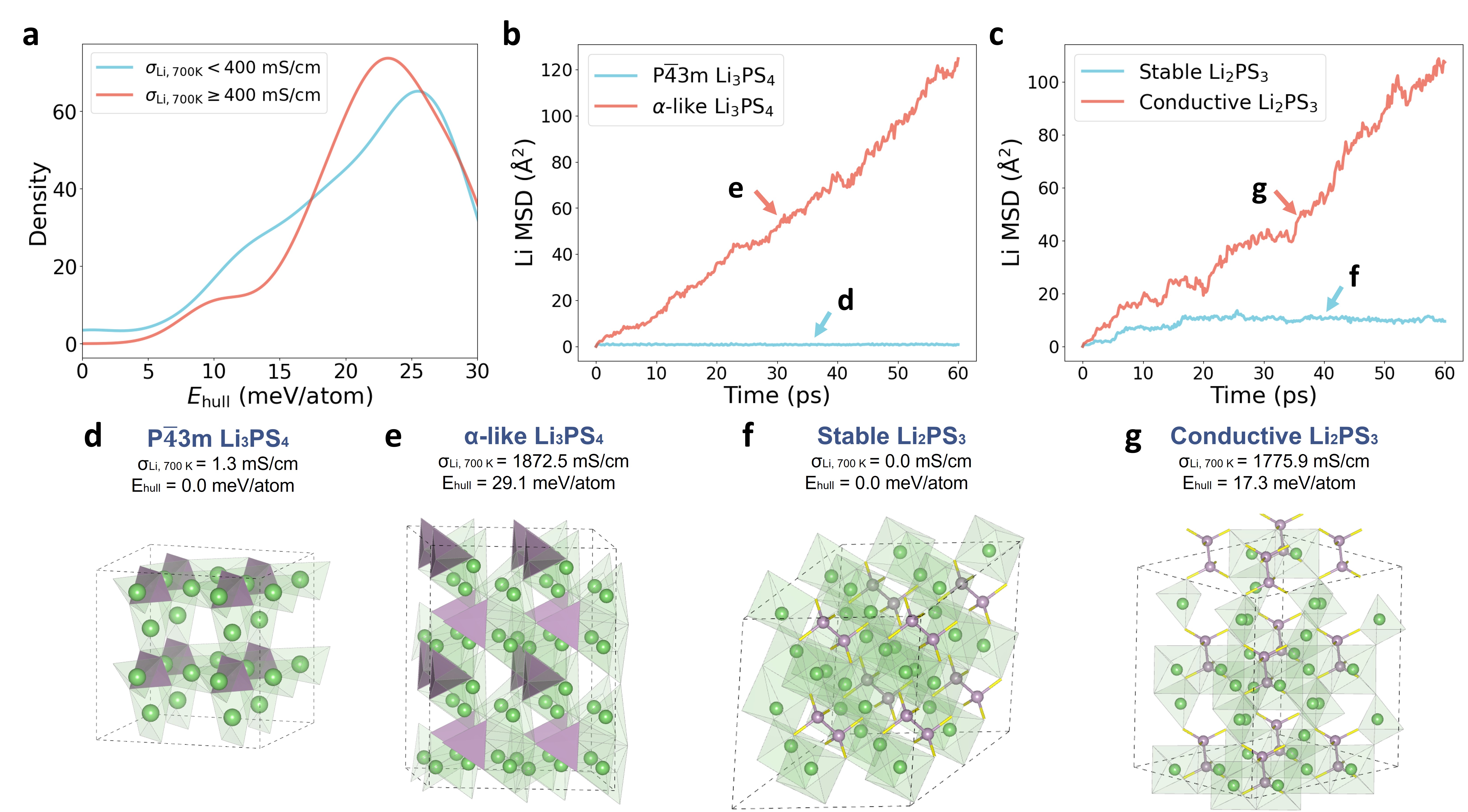}
    \caption{Ionic-transport screening results. (a) Distribution densities of kinetically stable Li-ion conductors (red, $\sigma_{\mathrm{Li}} \geq$ 400 mS/cm) and low-conductivity structures (cyan, $\sigma_{\mathrm{Li}} <$ 400 mS/cm) over $E_{\mathrm{hull}}$. (b-c) Li MSD at 700 K as a function of simulation time in example (b) Li$_3$PS$_4$ and (c) Li$_2$PS$_3$ polymorphs. Red curves indicate the metastable conductive polymorph, whereas cyan curves represent the most stable version. (d-g) Crystal structures of the example polymorphs. (d) P$\bar{4}$3m Li$_3$PS$_4$. (e) $\alpha$-like Li$_3$PS$_4$. (f) Stable Li$_2$PS$_3$. (g) Conductive Li$_2$PS$_3$. These structures were identified through iterative exploration and are absent from the MP database.}
    \label{fig:fig6}
\end{figure*}

Using the generated structures, we performed high-throughput molecular dynamics (MD) simulations at 700 K for 60 ps to screen Li-ion mobility among iteratively generated structures. Prior to MD, structures were filtered according to composition and stability criteria: $x_{\mathrm{Li}}>0.15$, $x_{\mathrm{P}}>0$, $x_{\mathrm{S}}>0.45$, $E_{\mathrm{hull}}\leq0.03$ eV/atom, and specific volume per Li atom $<200\,\mathrm{\AA}^3$. To exclude kinetically unstable structures, we further required the diffusivities of P and S to remain below $10^{-6}$ cm$^2$/s. Candidate fast-ion conductors were identified using a Li-ion conductivity threshold of 400 mS/cm, with $\sigma_{\mathrm{Li}} = n_{\mathrm{Li}} q^2 D_{\mathrm{Li}}/(k_\mathrm{B}T)$.

This screening identified 29 candidate Li-ion conductors absent from the MP database (Appendix Table~\ref{tab:tab-s1} and Appendix Figure~\ref{fig:figs5}). As shown in Figure~\ref{fig:fig6}a, the conductive candidates tend to lie at slightly higher $E_{\mathrm{hull}}$ than the non-conducting structures, consistent with the empirical observation that moderate metastability can promote local structural fluctuations and low-barrier Li$^+$ migration pathways\cite{junDiffusionMechanismsFast2024}.
Figure~\ref{fig:fig6}b compares Li mean-squared displacements (MSDs) in the on-hull P$\bar{4}$3m Li$_3$PS$_4$ phase (Figure~\ref{fig:fig6}d) and a metastable $\alpha$-like Li$_3$PS$_4$ polymorph\cite{hommaCrystalStructurePhase2011} (Figure~\ref{fig:fig6}e). In the P$\bar{4}$3m phase, the PS$_4^{3-}$ tetrahedra are arranged in a highly symmetric environment that stabilizes Li sites and suppresses diffusion. The $\alpha$-like polymorph exhibits lower local symmetry and enhanced Li mobility. A similar contrast is observed for Li$_2$PS$_3$ (Figure~\ref{fig:fig6}c,f,g), where the most stable phase confines Li to octahedral sites, whereas the more conductive polymorph additionally accommodates tetrahedrally coordinated Li.

Some predicted high-mobility candidates, including Li$_2$PS$_3$, have been reported experimentally to exhibit limited ionic transport under ambient conditions\cite{dietrichLocalStructuralInvestigations2016}. Because the present simulations are performed at 700 K over 60 ps, the resulting conductivities should be interpreted as qualitative indicators of intrinsic lattice mobility rather than quantitative room-temperature transport values. These calculations demonstrate the use of the explored Li--P--S chemical space for structure-resolved screening of Li-ion mobility.

\subsection{Downstream applications: Electronic-structure prediction}

\begin{figure*}
    \centering
    \includegraphics[width=0.9\linewidth]{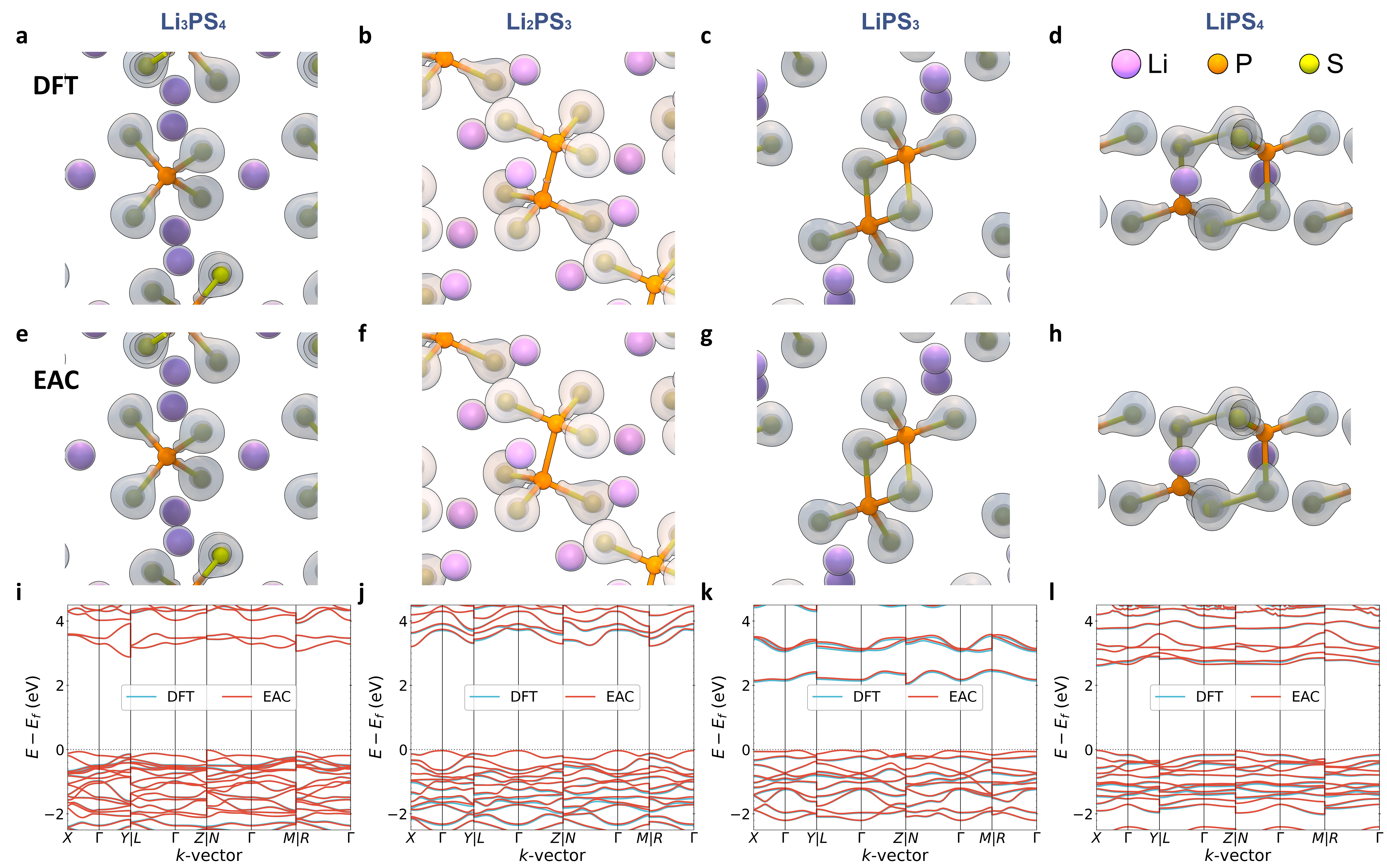}
    \caption{Electronic-structure prediction results. (a–h) Comparison of electronic charge densities for representative structures: (a, e) Li$_3$PS$_4$, (b, f) Li$_2$PS$_3$, (c, g) LiPS$_3$, and (d, h) LiPS$_4$. Panels a–d show charge densities obtained from self-consistent DFT calculations, while panels e–h display those predicted by the fine-tuned EAC-Net. (i–l) Electronic band-structure dispersions of the aforementioned representative compounds: (i) Li$_3$PS$_4$, (j) Li$_2$PS$_3$, (k) LiPS$_3$, and (l) LiPS$_4$. Cyan lines denote the DFT self-consistent (SCF) results, and red lines indicate non-self-consistent (NSCF) band structures computed using EAC-Net–predicted charge densities as input.}
    \label{fig:fig7}
\end{figure*}

As a demonstration beyond atomistic energetics, we further adapted an EAC-Net model\cite{xuejianEACNetRealspaceCharge2025} to the explored Li--P--S chemical space for electronic-structure prediction. The fine-tuned model was used to predict electronic charge densities, which were subsequently employed for non-self-consistent band-structure calculations.

Figures~\ref{fig:fig7}a–d and e–h compare DFT-computed charge densities with EAC-Net predictions for representative Li--P--S structures. The predicted densities show close agreement with the DFT references. Quantitatively, the normalized mean absolute error (NMAE) decreases from $\sim7.2\times10^{-3}$ at zero-shot inference to $\sim4.8\times10^{-3}$ after fine-tuning on only 34 self-consistent-field (SCF) configurations (Appendix Figure~\ref{fig:figs6}e), indicating efficient adaptation across the sampled Li--P--S compositions and bonding environments.

When the ML-predicted charge densities are used as inputs for non-self-consistent (NSCF) band-structure calculations, the resulting dispersions closely reproduce those obtained from DFT SCF calculations (Figure~\ref{fig:fig7}i–l), including both the overall band-gap characteristics and finer features near the Fermi level. These results show that the chemical-system–adapted representation can be extended from atomistic energetics to electronic-structure prediction with limited additional first-principles data.

\section{Discussion}\label{sec:dis}

In this work, we demonstrated a chemical-system–centric strategy for crystalline materials exploration using the Li--P--S ternary system as a test case. By coupling generative modeling, MLFF-based energetic evaluation, selective DFT labeling, and iterative model refinement, the workflow generated approximately 70,000 candidate structures over seven exploration cycles, including more than 10,000 stable–unique–novel structures. The diversity of short-range local environments saturated within the first few iterations, while MLFF prediction errors decreased concurrently. The exploration also recovered chemically plausible P–S motifs absent from the training databases but consistent with earlier experimental reports. The resulting system-adapted models further supported finite-$P$--$T$ phase stability calculations, Li-ion transport screening, and electronic-structure prediction within the same chemical space. 

These results suggest that operational saturation of near-equilibrium configurations and short-range atomic environments was observed within a bounded chemical system, while additional structural diversity can largely arise from different spatial arrangements, topological combinations, and long-range ordering of already sampled local motifs. Recovery of chemically plausible local motifs further suggests that generative models coupled with energetic constraints from MLFFs can, to some extent, generalize learned chemical bonding preferences beyond the specific structures present in the training data, enabling exploration of a broader range of chemically plausible configurations. In this context, generative models should not be evaluated solely by their ability to reproduce individual crystal structures, but more importantly by their capability to efficiently propose diverse and chemically meaningful configurations, as this capability determines the breadth of configurational space that can be explored within an iterative discovery loop.

This strategy should be viewed as complementary to, rather than a replacement for, universal atomistic models. Universal foundation models provide broad chemical priors across many systems, whereas bounded chemical-system refinement can concentrate model capacity on a restricted domain, reduce the cost of repeated high-throughput evaluation, and provide an explicit convergence criterion for local chemical coverage. A practical hierarchy may therefore combine universal pretrained models with subsequent system-specific exploration and adaptation, while the resulting domain-specific data can in turn expand the training sets of universal models.

Several limitations remain in the present study. The demonstration is limited to the Li--P--S system; although it is chemically complex, different chemistries may exhibit different saturation behavior and first-principles data requirements. In addition, the chemical-space completeness discussed here is restricted to near-equilibrium stable and metastable crystalline configurations and to short-range environments resolved by the employed local descriptors and models. It does not imply coverage of transition states, rare-event pathways, strongly nonequilibrium configurations, or exhaustive long-range ordering and supercell-scale arrangements, which may require incorporating dedicated sampling methods into the data-generation framework. Therefore, local structural and energetic completeness does not guarantee experimental synthesizability of the discovered structures, which depends on kinetic pathways, precursors, processing conditions, and other factors beyond the present framework. Looking forward, extending this chemical-system–centric strategy across diverse chemistries, while incorporating specialized sampling methods for kinetically and structurally complex regimes, may provide a practical route toward scalable and progressively more complete computational materials exploration.

\acknowledgments
This work was supported by the National Key R\&D Program of China (Grants No. 2021YFA0718900), National Natural Science Foundation of China (Grants No. 92477114 and No. 12374096), and the Suzhou Municipal Science and Technology Program (Gusu Innovation and Entrepreneurship Leading Talent Program, Grant No. ZXL2025304). We thank DP Technology for providing computational resources through the Bohrium platform and data hosting services via AIS Square.

\begin{figure*}[!t]
    \centering
    \includegraphics[width=0.9\linewidth]{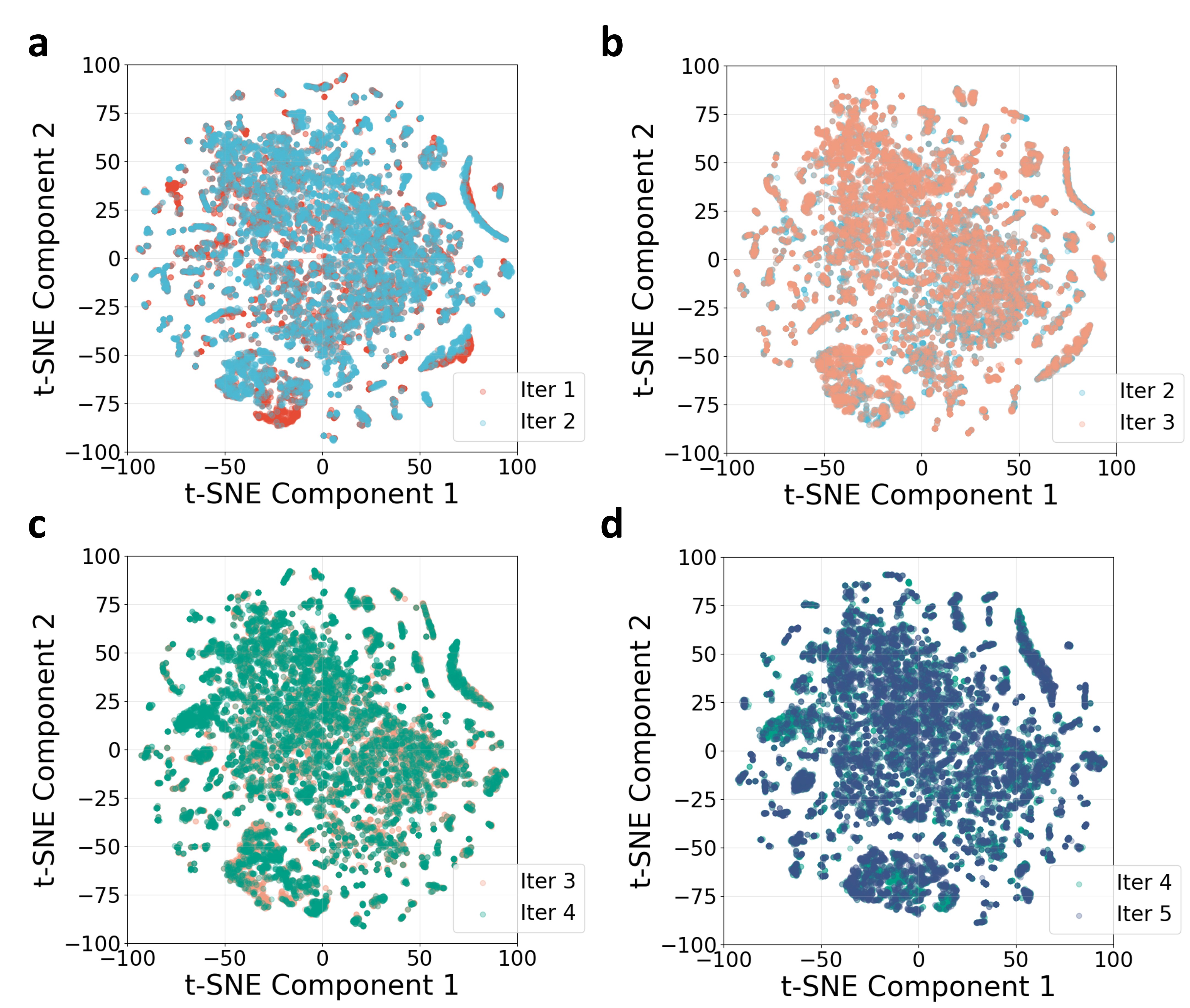}
    \caption{Comparison of the first two t-SNE components of structural fingerprints between (a) iterations 1 and 2, (b) iterations 2 and 3, (c) iterations 3 and 4, and (d) iterations 4 and 5. The t-SNE features were computed by jointly embedding all generated structures across iterations to ensure alignment, and then plotted every two iterations for comparison.}
    \label{fig:figs2}
\end{figure*}

\appendix
\section{Local-environment and anion-motif analysis}
\subsection{Structural fingerprints}
Local structural fingerprints were computed following the same procedure as in CDVAE\cite{xieCrystalDiffusionVariational2021}. Each atomic site was featurized using the CrystalNNFeaturizer (``ops'' preset) implemented in matminer\cite{wardMatminerOpenSource2018}, with oxidation states assigned via pymatgen’s built-in valence-guessing method\cite{ongPythonMaterialsGenomics2013}. A structure-level fingerprint was then obtained by averaging all site vectors over each dimension. For analyses focusing on the anionic framework, all Li atoms were removed from the structures before recomputing fingerprints, yielding the t-SNE projection presented in Figure~\ref{fig:fig4}a of the main text.

To further examine the evolution of local structural environments across successive exploration cycles, we jointly embedded the structural fingerprints from all iterations into the same t-SNE space. Pairwise comparisons between consecutive iterations are shown in Fig.~\ref{fig:figs2}. The distributions become strongly overlapping after the early iterations, consistent with the saturation of local-environment diversity discussed in Section~\ref{sec:saturation}.

\subsection{Anion motif extraction and classification}

To identify and categorize distinct thiophosphate anionic motifs within generated Li--P--S structures, we developed a graph-based motif extraction pipeline implemented with pymatgen\cite{ongPythonMaterialsGenomics2013}, ASE\cite{larsenAtomicSimulationEnvironment2017}, and NetworkX\cite{hagbergExploringNetworkStructure2008}.

Each structure was first converted into a connectivity graph where nodes represent atomic sites and edges represent chemical bonds between P and S atoms (Li atoms were excluded). Interatomic distances were compared against empirical cutoffs of 2.35 $\mathrm{\AA}$ (P–S), 2.20 $\mathrm{\AA}$ (S–S), and 2.50 $\mathrm{\AA}$ (P–P). An undirected graph was constructed using these criteria, yielding a P–S bonding network for each structure. The bond graph was then decomposed into connected components using the NetworkX connected-component algorithm, where each subgraph corresponds to a unique anionic cluster such as PS$_4^{3-}$ or P$_2$S$_6^{4-}$.

To distinguish topologically distinct motifs, each subgraph was encoded using the Weisfeiler–Lehman (WL) graph hash\cite{shervashidzeWeisfeilerLehmanGraphKernels2011} method, with atomic species as node attributes and bond type (P–S, S–S, P–P) as edge attributes. This ensured an isomorphism-invariant representation, allowing unique identification of anion topologies. Each isolated unique cluster was then re-centered in a cubic box (20 $\mathrm{\AA}$) and exported as an ASE Atoms object. All skeletal formulae appearing in Figure~\ref{fig:fig4} were visualized using a demo version of the Marvin chemical drawing tool\cite{MarvinChemicalDrawing}.

\begin{figure}[!htbp]
    \centering
    \includegraphics[width=0.9\linewidth]{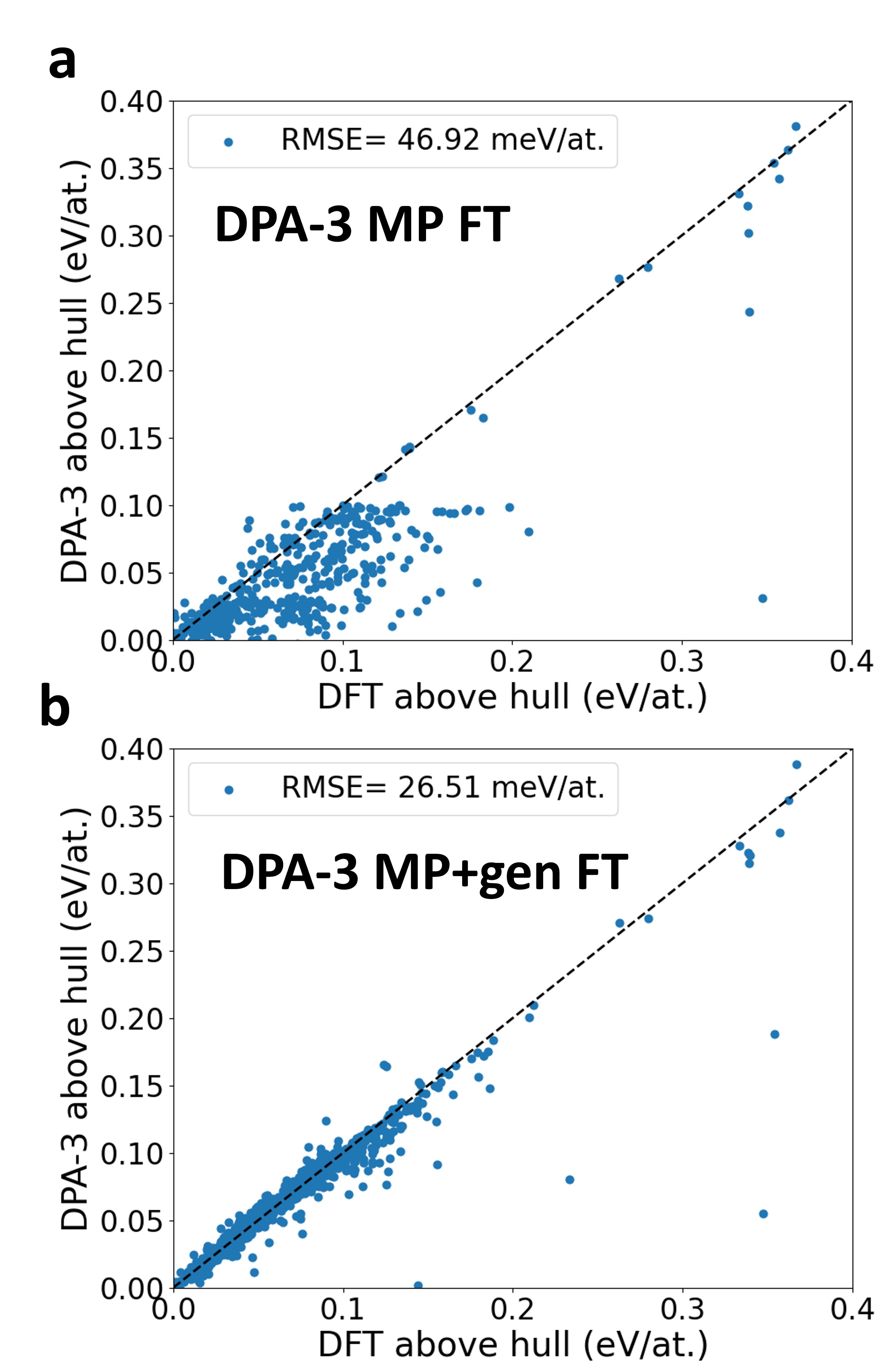}
    \caption{Parity plots comparing DPA-3 predicted energies above hull with DFT-computed values: (a) fine-tuned using only Li--P--S structures from the MP dataset, and (b) fine-tuned using combined MP and generated data at $N_{\mathrm{train}}=4050$.}
    \label{fig:figs3}
\end{figure}

\begin{figure*}[!t]
    \centering
    \includegraphics[width=1.0\linewidth]{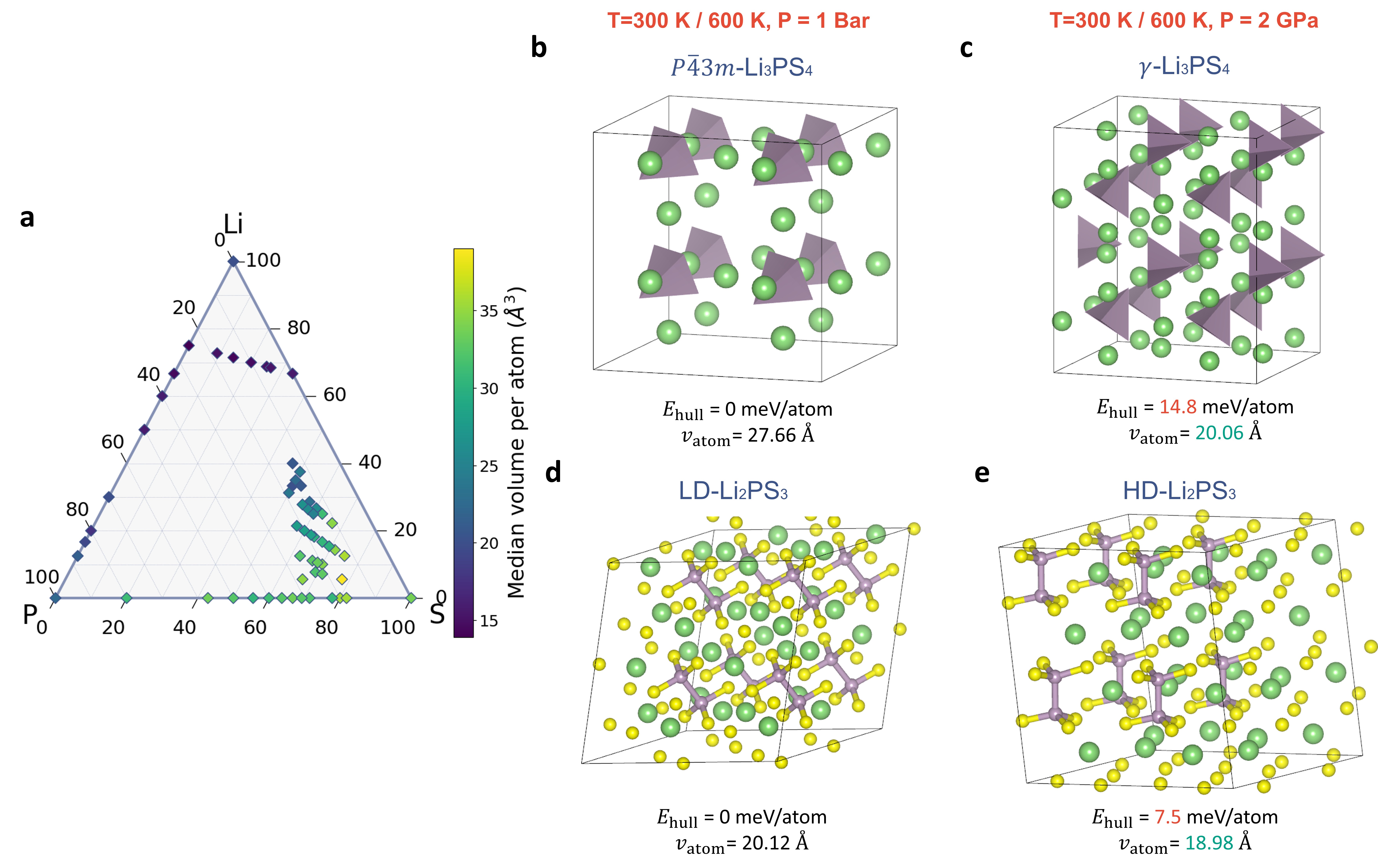}
    \caption{(a) Median specific volume per atom as a function of composition in the Li--P--S ternary system. Darker blue dots indicate denser atomic packing. (b–c) Lowest Gibbs free-energy structures of Li$_3$PS$_4$ at $T=300$–$600$ K under (b) 1 bar (high-symmetry P$\bar{4}$3m phase) and (c) 2 GPa (the $\gamma$ phase\cite{hommaCrystalStructurePhase2011}). (d–e) Lowest Gibbs free-energy structures of Li$_2$PS$_3$ at $T=300$–$600$ K under (d) 1 bar (low-density, LD phase) and (e) 2 GPa (high-density, HD phase).}
    \label{fig:figs4}
\end{figure*}

\section{Additional validation of the adapted MLFF}
In addition to the energy and force errors reported in the main text, we evaluated the accuracy of DPA-3 for predicting \(E_{\mathrm{hull}}\), which directly controls stability ranking during iterative exploration. Figure~\ref{fig:figs3} compares models fine-tuned using only Li--P--S structures from the MP dataset and using the combined MP and generated dataset. Inclusion of iteratively generated configurations substantially improves agreement with DFT across the sampled Li--P--S domain, demonstrating the importance of expanding the training set beyond the initial database inventory.

\begin{figure*}[!t]
    \centering
    \includegraphics[width=0.9\linewidth]{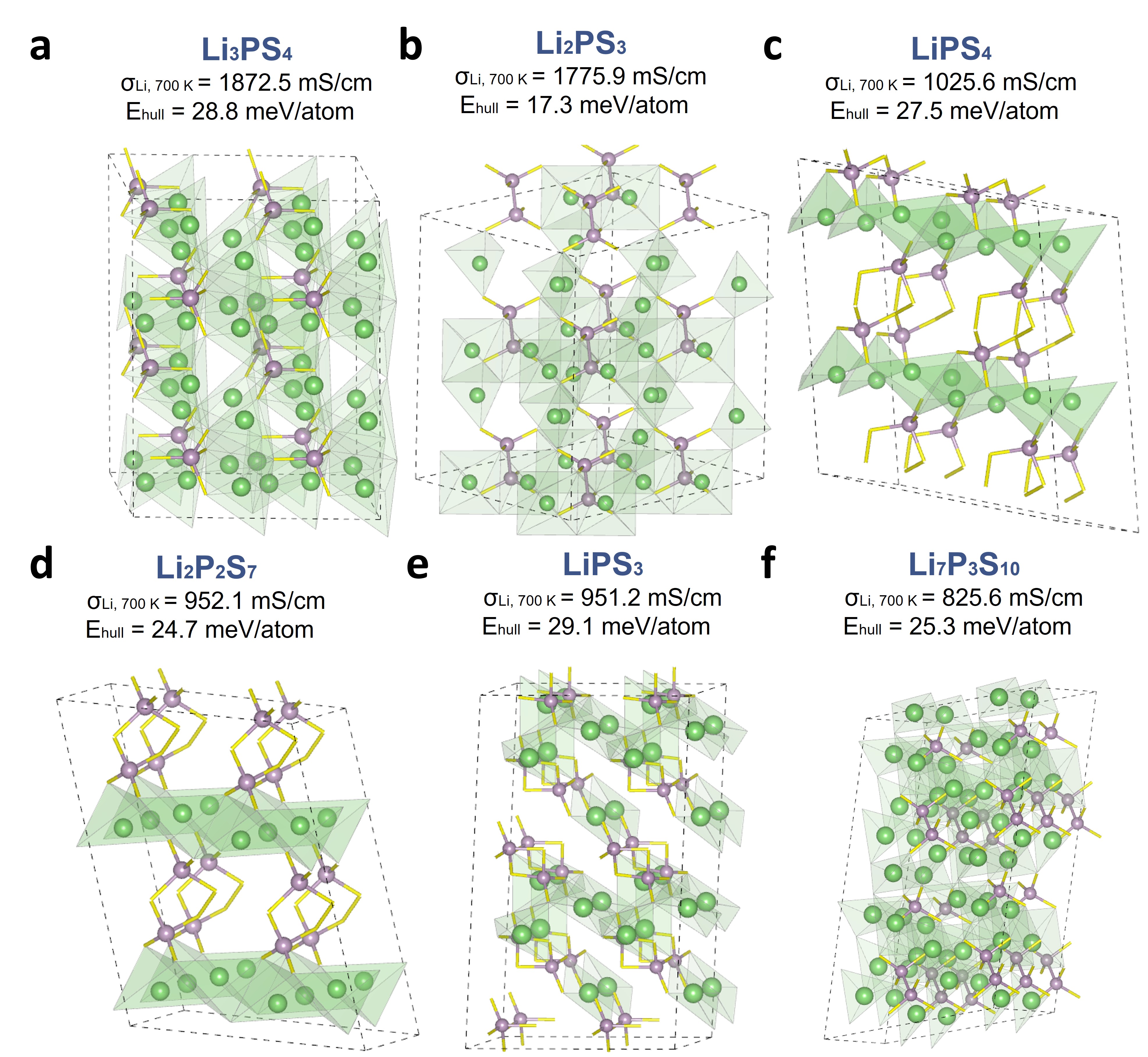}
    \caption{Crystal structures of representative Li-ion conductors listed in Table~\ref{tab:tab-s1}. Green spheres and transparent polyhedra denote Li ions and their coordination environments, respectively, purple spheres represent P atoms, and purple–yellow sticks indicate P–S bonds. S atoms are omitted for clarity.}
    \label{fig:figs5}
\end{figure*}

\section{Finite-$P$--$T$ phase-stability calculations}

Finite-pressure and finite-temperature phase diagrams were computed using the fine-tuned DPA-3 MLFF at $N_{\mathrm{train}}=4050$ as the energy and force evaluator. Structural perturbations and force constants were generated with phonopy\cite{togoImplementationStrategiesPhonopy2023, togoFirstprinciplesPhononCalculations2023}. To prevent spurious interactions between periodic images, all structures were scaled to ensure lattice parameters of at least $12\mathrm{\AA}$ and approximately equal lengths in each direction.

Within the quasi-harmonic approximation (QHA)\cite{togoFirstprinciplesPhononCalculations2010}, phonon frequencies were obtained by scaling lattice parameters from 0.85 × $a_0$ to 1.15 × $a_0$ in increments of 0.05. The phonon density of states and mode frequencies were then integrated to compute the Helmholtz free energy $F(T,V)$. Finally, the Gibbs free energy $G(T,p)$ was derived through Legendre transformation in the temperature range 0–600 K and pressure range 0–2 GPa. Phase diagrams at varied temperatures and pressures were then constructed by evaluating the Gibbs free energy convex hulls.

The composition dependence of the median specific volume is shown in Fig.~\ref{fig:figs4}a. S-rich compositions generally exhibit lower packing density, which contributes to their increased \(PV\) penalty at elevated pressure. Figures~\ref{fig:figs4}b–e further show the lowest-Gibbs-free-energy structures of Li$_3$PS$_4$ and Li$_2$PS$_3$ under representative thermodynamic conditions, illustrating the structural changes underlying the phase-stability trends discussed in the main text.

\section{Ionic-transport screening}

As a preliminary demonstration of the ionic-conductivity screening capability, we performed MD simulations to estimate Li-ion diffusivity at elevated temperature (700 K) over a short simulation time of 60 ps. The resulting conductivities therefore serve as qualitative indicators rather than quantitative room-temperature values.

MD simulations were carried out in ASE using the non-fine-tuned MatterSim-5M\cite{MatterSimPretrained_modelsMain} model, which was pretrained on high-temperature and high-pressure datasets and exhibits better numerical stability under such conditions. All simulation cells were scaled to ensure lattice parameters of at least 10 $\mathrm{\AA}$ and approximately isotropic dimensions.

Each system was equilibrated in the NPT ensemble at 1 bar for 10 ps using a Nosé–Hoover thermostat and a 2 fs timestep, followed by 60 ps of production simulation in the NVT ensemble for diffusivity analysis. Elemental diffusivities were obtained by fitting the mean-squared displacement (MSD) in the final 50 ps of the NVT trajectory according to
\begin{equation}
    D_{\mathrm{element}} = \frac{\left<||r_i(t_1) - r_i(t_0)||^2\right>_{i \in \mathrm{element}}}{6(t_1 - t_0)}.
\end{equation}
The Li-ion conductivity was then calculated from the Einstein relation
\begin{equation}
    \sigma_{\mathrm{Li}} = \frac{n_{\mathrm{Li}}z_{\mathrm{Li}}^2 e^2 D_{\mathrm{Li}}}{k_{\mathrm{B}}T},
\end{equation}
where $n_{\mathrm{Li}}$ is the Li atomic number density and $z_\mathrm{Li}=1$.

Applying the screening criteria described in the main text yielded 29 structures with high Li mobility at 700 K. Their compositions, specific volumes, predicted conductivities, and \(E_{\mathrm{hull}}\) values are summarized in Table~\ref{tab:tab-s1}. These values should be interpreted as qualitative screening metrics rather than quantitative room-temperature conductivities.

\begin{table*}[!t]
\begin{minipage}[t]{0.58\textwidth}\vspace{0pt}
\caption{Summary of Li-ion conductor candidates discovered by conductivity screening. $v_{\mathrm{atom}}$ represents specific volume per atom, $\sigma_{\mathrm{Li, 700\ K}}$ represents Li-ion conductivity at 700 K, and $E_{\mathrm{hull}}$ represents the energy above hull evaluated using the fine-tuned DPA-3. (*) Same anion arrangement as the $\alpha$-phase\cite{hommaCrystalStructurePhase2011}. (\#) $\beta$-structure, entry from MP database. (!) Li$_7$P$_3$S$_{11}$, entry from MP database.}
\label{tab:tab-s1}
\centering\footnotesize\setlength{\tabcolsep}{2pt}\renewcommand{\arraystretch}{1.08}
\begin{tabular}{@{}lllccc@{}}
Iteration & Name & Formula & \shortstack{$v_{\mathrm{atom}}$\\($\mathrm{\AA}^3$/atom)} & \shortstack{$\sigma_{\mathrm{Li, 700\ K}}$\\(mS/cm)} & \shortstack{$E_{\mathrm{hull}}$\\(meV/atom)} \\
\midrule
3 & ea\_0.00\_gen\_2866(*)&  Li$_3$PS$_4$ & 21.8 & 1872.5 & 28.8 \\
2 & ea\_0.00\_gen\_1016 &  Li$_2$PS$_3$ & 21.2 & 1775.9 & 17.3 \\
1 & ea\_0.03\_gen\_2999 &  Li$_2$PS$_3$ & 21.4 & 1490.2 & 10.7 \\
0 & mp-985583-GGA(\#) &  Li$_3$PS$_4$ & 21.8 & 1397.6 & 20.6 \\
1 & ea\_0.00\_gen\_2765(*) &  Li$_3$PS$_4$ & 22.1 & 1282.8 & 24.3 \\
2 & ea\_0.00\_gen\_1404 &  Li$_3$PS$_4$ & 23.1 & 1198.2 & 29.6 \\
3 & ea\_0.06\_gen\_823  &  Li$_3$PS$_4$ & 26.1 & 1163.1 & 23.3 \\
2 & ea\_0.00\_gen\_1607(*) &  Li$_3$PS$_4$ & 21.9 & 1142.8 & 20.0 \\
0 & mp-641703-GGA(!) &  Li$_7$P$_3$S$_{11}$ & 22.4 & 1059.9 & 28.9 \\
5 & ea\_0.03\_gen\_2294 &  LiPS$_4$ & 32.1 & 1025.6 & 27.5 \\
1 & ea\_0.00\_gen\_1603 &  Li$_2$P$_2$S$_7$ & 31.5 &  952.1 & 24.7 \\
3 & ea\_0.03\_gen\_1068 &  LiPS$_3$     & 28.0 &  951.2 & 29.0 \\
1 & ea\_0.06\_gen\_410  &  Li$_2$PS$_3$ & 20.8 &  921.1 & 19.9 \\
1 & ea\_0.03\_gen\_3012 &  Li$_2$PS$_3$ & 24.1 &  826.6 & 21.4 \\
1 & ea\_0.03\_gen\_1288 &  Li$_7$P$_3$S$_{10}$ & 22.1 &  825.6 & 25.3 \\
2 & ea\_0.00\_gen\_468  &  LiPS$_3$     & 27.7 &  819.6 & 22.6 \\
3 & ea\_0.00\_gen\_371  &  LiPS$_3$     & 30.5 &  758.5 & 24.2 \\
1 & ea\_0.00\_gen\_270  &  LiPS$_3$     & 26.7 &  745.9 & 24.0 \\
1 & ea\_0.03\_gen\_701  &  Li$_2$PS$_3$ & 20.7 &  682.2 & 19.6 \\
3 & ea\_0.06\_gen\_2550 &  LiPS$_4$     & 30.7 &  671.6 & 23.2 \\
1 & ea\_0.00\_gen\_1470 &  Li$_2$PS$_3$ & 20.9 &  625.7 & 9.6 \\
5 & ea\_0.06\_gen\_1312 &  LiPS$_4$ & 40.0 &  575.1 & 28.1 \\
1 & ea\_0.06\_gen\_1779 &  Li$_2$PS$_3$ & 20.6 &  498.7 & 23.3 \\
3 & ea\_0.03\_gen\_1953 &  Li$_2$P$_2$S$_7$ & 37.5 &  471.2 & 28.2 \\
4 & ea\_0.06\_gen\_1317 &  LiPS$_4$ & 34.0 &  462.2 & 21.4 \\
2 & ea\_0.00\_gen\_538  &  Li$_2$P$_2$S$_7$ & 34.2 &  444.7 & 25.0 \\
1 & ea\_0.00\_gen\_2221 &  LiPS$_4$ & 31.2 &  443.0 & 29.9 \\
1 & ea\_0.03\_gen\_420 &  LiPS$_3$ & 26.9 &  436.1 & 17.4 \\
7 & ea\_0.00\_gen\_493  &  LiPS$_4$ & 34.5 &  435.4 & 15.2 \\
7 & ea\_0.03\_gen\_3148 &  LiPS$_3$ & 31.7 &  428.8 & 18.7 \\
6 & ea\_0.00\_gen\_1281 &  LiPS$_4$ & 36.5 &  406.4 & 21.3 \\
\bottomrule
\end{tabular}
\end{minipage}
\hfill
\begin{minipage}[t]{0.40\textwidth}\vspace{0pt}\figurecaptiontype
    \centering
    \includegraphics[width=\linewidth]{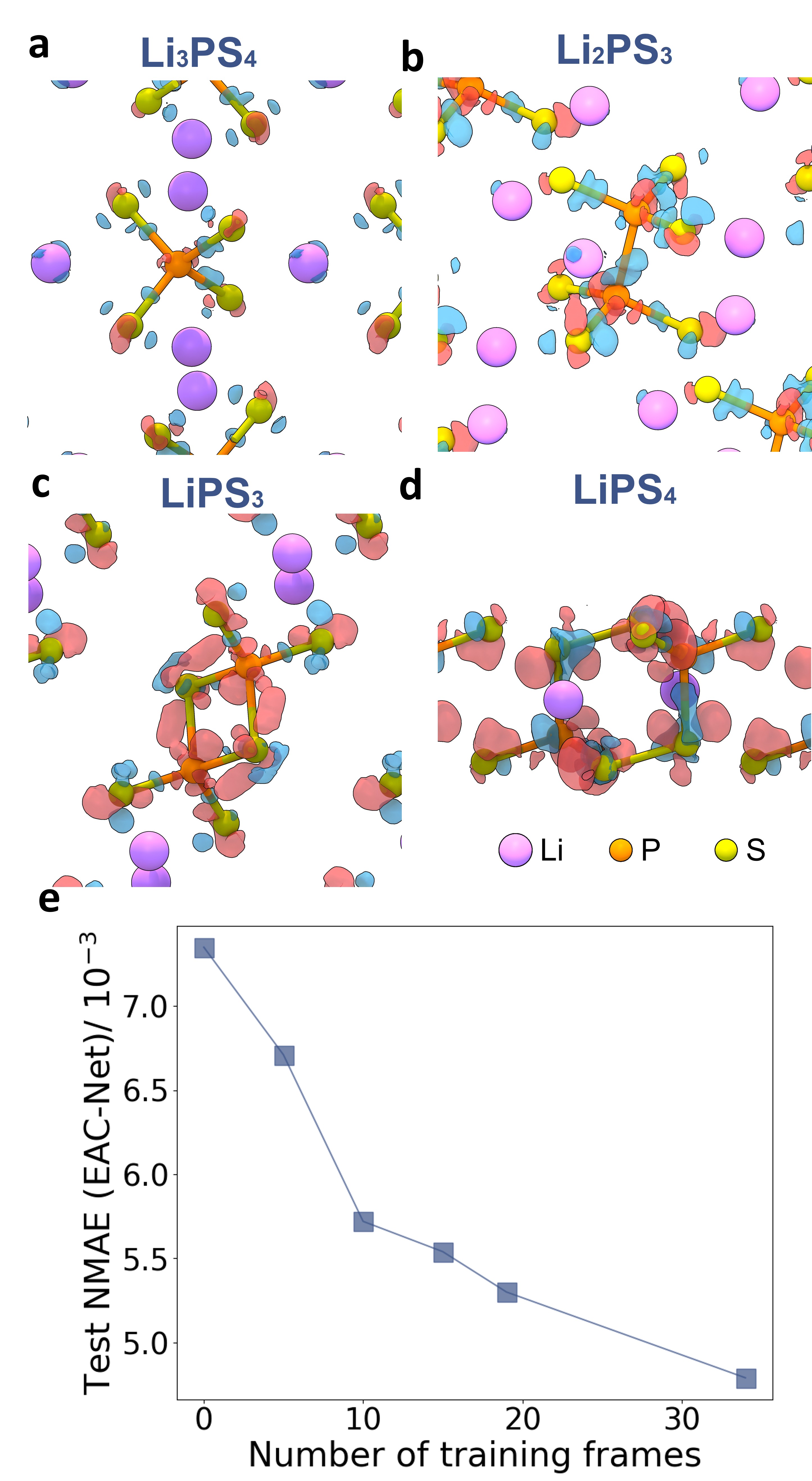}
    \caption{(a–d) Differences between charge densities predicted by EAC-Net and those computed by DFT for (a) Li$_3$PS$_4$, (b) Li$_2$PS$_3$, (c) LiPS$_3$, and (d) LiPS$_4$, corresponding to the order in Figure~7 of the main text. (e) Normalized mean absolute error (NMAE) of the predicted charge density as a function of the number of fine-tuning frames.}
    \label{fig:figs6}
\end{minipage}
\end{table*}

Representative structures passing the stability and kinetic filters are shown in Fig.~\ref{fig:figs5}.

\section{Electronic structure prediction}
The Equivariant Atomic Contribution Network (EAC-Net) was employed to predict electronic charge densities. Specifically, we adopted the pretrained EAC-mp model developed in previous work\cite{xuejianEACNetRealspaceCharge2025}, which was trained on 48,183 CHGCAR files randomly sampled from the Materials Project (MP) database\cite{jainCommentaryMaterialsProject2013}. The model utilized an angular momentum cutoff of $l_{\mathrm{max}} = 5$ and a feature dimension of 36. The cutoff radii for atomic and grid representations were set to 4.0 $\mathrm{\AA}$ and 6.0 $\mathrm{\AA}$, respectively, yielding a total of approximately 3.08 million trainable parameters. Comprehensive details on dataset composition, model architecture, training procedures, and error quantification can be found in the original EAC-Net publication.

\newpage
For fine-tuning, we selected 5, 10, 15, 19, and 34 LPS frames from iterations 1 and 2 to evaluate convergence of the model loss with respect to the number of fine-tuning configurations. Each model was trained for 2000 steps with a batch size of 50, using a learning rate decaying monotonically from $1\times10^{-4}$ to $8\times10^{-6}$. Ten test configurations were selected from iteration 6 for validation.

\newpage
Self-consistent-field (SCF) and non-self-consistent-field (NSCF) DFT calculations were performed using the same computational parameters as those employed for MLFF dataset generation, except that the electronic convergence threshold was tightened to $10^{-5}$ eV, and the KSPACING parameter was set to $0.2\,\mathrm{\AA}^{-1}$ to ensure the accuracy of the charge-density results.

\newpage
Figure~\ref{fig:figs6}a–d shows the spatial errors between EAC-Net-predicted and DFT charge densities for the representative structures discussed in the main text. The remaining errors are localized primarily around atomic regions, while the overall density distribution is well reproduced. Figure~\ref{fig:figs6}e shows the convergence of NMAE with the number of fine-tuning configurations, supporting the use of 34 configurations in the final model.


\clearpage

\let\originalunderscore\_
\renewcommand{\_}{\originalunderscore\allowbreak}
\providecommand{\bibinfo}[2]{#2}
\let\originalbibinfo\bibinfo
\renewcommand{\bibinfo}[2]{%
  \def\bibfieldname{#1}\def\urlfieldname{howpublished}%
  \ifx\bibfieldname\urlfieldname
    \layoutbiburl{#2}%
  \else
    \originalbibinfo{#1}{#2}%
  \fi
}
\bibliography{refs}

\end{document}